\documentclass[twocolumn,showpacs,preprintnumbers,amsmath,amssymb]{revtex4}
\usepackage{graphicx}
\usepackage{dcolumn}
\usepackage{bm}

\begin{document}

\preprint{\today}

\title{Electrostatic trapping of metastable NH molecules}
\author{Steven Hoekstra}
\email{hoekstra@fhi-berlin.mpg.de}
\author{Markus Mets\"al\"a}
\author{Peter C.~Zieger}
\author{Ludwig Scharfenberg}
\author{Joop J.~Gilijamse}
\author{Gerard Meijer}
\author{Sebastiaan Y.~T.~van de Meerakker}
\affiliation{Fritz-Haber-Institut der Max-Planck-Gesellschaft,
Faradayweg 4-6, 14195 Berlin, Germany}

\begin{abstract}
We report on the Stark deceleration and electrostatic trapping of
$^{14}$NH ($a\,^1\Delta$) radicals. In the trap, the molecules are
excited on the spin-forbidden $A\,^3\Pi \leftarrow a\,^1\Delta$
transition and detected via their subsequent fluorescence to the
$X\,^3\Sigma^-$ ground state. The 1/e trapping time is 1.4 $\pm$
0.1 s, from which a lower limit of 2.7 s for the radiative
lifetime of the $a\,^1\Delta, v=0,J=2$ state is deduced. The
spectral profile of the molecules in the trapping field is
measured to probe their spatial distribution. Electrostatic
trapping of metastable NH followed by optical pumping of the
trapped molecules to the electronic ground state is an important
step towards accumulation of these radicals in a magnetic
trap.\end{abstract}

 \maketitle
\section{Introduction}

The ability to confine cold gas-phase (polar) molecules in traps
has increased interest in gas-phase molecular physics in recent
years. This interest stems from the exotic and intriguing
phenomena that are predicted to be observable in these cold
molecular gases, ranging from chemistry at ultralow
temperatures~\cite{Krems:2005fk} to the engineering of macroscopic
collective quantum states~\cite{Baranov2002}. The field of
(ultra)cold molecules might revolutionize molecular physics. It
even has the potential to affect physics at large, just as the
first successful production of (ultra)cold atoms has led to
spectacular advances far beyond the realm of classical atomic
physics. The techniques to produce cold molecules are slowly
maturing; photo-association~\cite{Takekoshi1998}, buffer gas
cooling~\cite{WEINSTEIN1998}, and Stark
deceleration~\cite{BETHLEM2000} have already resulted in the three
dimensional trapping of molecules. Many other techniques are
currently being developed by various groups.

The method of Stark deceleration and electrostatic trapping
combines molecular beam technology with concepts from charged
particle accelerator physics. In essence, a part of a molecular
beam pulse of polar neutral molecules can be decelerated to a
standstill and confined in a trap, exploiting the interaction of
polar molecules with inhomogeneous electric fields. The technique
works for a large class of polar molecules, and has resulted to
date in the trapping of ND$_3$ molecules~\cite{BETHLEM2000}, OH
radicals~\cite{VANDEMEERAKKER2005,sawyer2007} and metastable CO
molecules~\cite{Gilijamse2007}. The density of the decelerated and
trapped molecular packets has proven to be sufficient for a number
of experiments~\cite{VANVELDHOVEN2004, Hudson2005B, GILIJAMSE2006,
VANDEMEERAKKER2005A,Hoekstra:2007ly}. However, it is still a
formidable challenge to reach high enough densities in the trap
that collisions occur between the trapped molecules on the
trapping time scale. This is required to be able to experimentally
study the interactions between the trapped molecules, and it is a
prerequisite for the future application of cooling schemes like
evaporative or sympathetic cooling.

An obvious route towards higher number densities in the trap is
the accumulation of several packets of molecules that are produced
in distinct cycles of the experiment. Although an increase of the
phase-space density in this way is forbidden by Liouville's
theorem, a reloading scheme has been proposed for the NH radical
that circumvents this fundamental
obstacle~\cite{VANDEMEERAKKER2001}. In this scheme, NH radicals
that are produced in the long-lived metastable $a\,^1\Delta$ state
are Stark decelerated to a standstill and subsequently optically
excited to the $A\,^{3}\Pi$ state. Spontaneous fluorescence from
the $A\,^{3}\Pi$ state to the $X\,^{3}\Sigma^{-}$ ground state
provides the required uni-directional pathway. A magnetic
quadrupole trap can be placed at the end of the Stark decelerator,
such that the created ground state molecules are magnetically
trapped. The Stark interaction in the ground state is rather weak,
therefore the Stark decelerator can be used to decelerate the next
pulse of metastable NH radicals without affecting the trapping
potential for ground state NH molecules. A number of prerequisites
for the successful implementation of this accumulation scheme have
already been experimentally demonstrated. The hitherto unobserved
$A\,^3\Pi \leftarrow a\,^1\Delta$ transition was found and
characterized \cite{VANDEMEERAKKER2003}, and in a preliminary
deceleration experiment, a beam of metastable NH radicals was
Stark decelerated from 550 m/s to 330
m/s~\cite{Vandemeerakker2006B}. However, the intensity of the
decelerated packet was rather poor, and prevented the
demonstration of deceleration to lower velocities.

Here we present the deceleration and electrostatic trapping of a
pulsed beam of NH ($a\,^1\Delta$) radicals. The trapping time of
the metastable radicals in the quadrupole electrostratic trap is
measured. A laser-based diagnostic method has been used to
characterize the spatial distribution of the trapped cloud.

The NH radical has several additional properties that make this
species of interest for cold molecule experiments. In fact, NH in
the $a\,^1\Delta$ state is one of the prime candidates for the
Stark deceleration and trapping technique. The Stark shift over
mass ratio (the parameter that determines the efficiency of the
deceleration process) is significantly larger than for the OH
radical, a system that is extensively used in deceleration and
trapping experiments. In the $X\,^3\Sigma^-$ ground state, its 2
$\mu_B$ magnetic moment makes the molecule a good candidate for
magnetic trapping experiments. Recently, ground state NH molecules
have been magnetically trapped using the buffer gas cooling
technique~\cite{Campbell:2007zr}, taking advantage of the
favorable properties of He-NH collisions~\cite{Krems2003B}. The
electric dipole allowed $A\,^3\Pi, v'=0 \leftarrow X\,^3\Sigma^-,
v''=0$ transition has an unusually large Franck-Condon
factor~\cite{YARKONY1989}, which offers interesting prospects for
direct laser cooling. It has also been predicted that NH in the
ground state has favorable properties to perform evaporative
cooling~\cite{Kajita2006B}. Finally, interesting collision
properties are expected between cold NH ($X\,^3\Sigma^-$)
molecules and ultracold Rb atoms~\cite{Soldan2004, Tacconi2007}.

\section{Experiment}

\begin{figure}
    \resizebox{\linewidth}{!}{\includegraphics{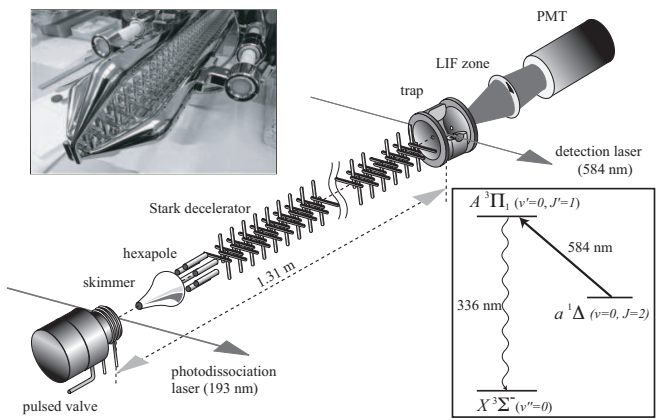}}
    \caption{\label{fig:setup} Scheme of the experimental setup. NH molecules in the $a^{1}\Delta$ state are created by photodissociation of HN$_{3}$ seeded in Kr. The beam passes through a skimmer, a hexapole, and a Stark decelerator. The decelerated molecules are subsequently loaded into an electrostatic trap. The molecules are probed at the center of the trap by a far off-resonant laser-induced fluorescence detection scheme, that is shown in the inset. A photograph of the Stark decelerator is shown in the upper left corner.}
    \end{figure}

The experiments to decelerate the NH radicals, and the experiments
to subsequently confine the molecules in an electrostatic
quadrupole trap, are performed in two different pulsed molecular
beam machines, operating at a repetition frequency of 10 Hz. The
deceleration experiments are performed in a newly constructed
Stark deceleration molecular beam machine; for the trapping
experiments the same setup is used as in our OH trapping
experiments. The latter setup is schematically depicted in
Figure~\ref{fig:setup}, and has been described in detail
elsewhere~\cite{VANDEMEERAKKER2005}.

\begin{figure}
\resizebox{\linewidth}{!}{\includegraphics{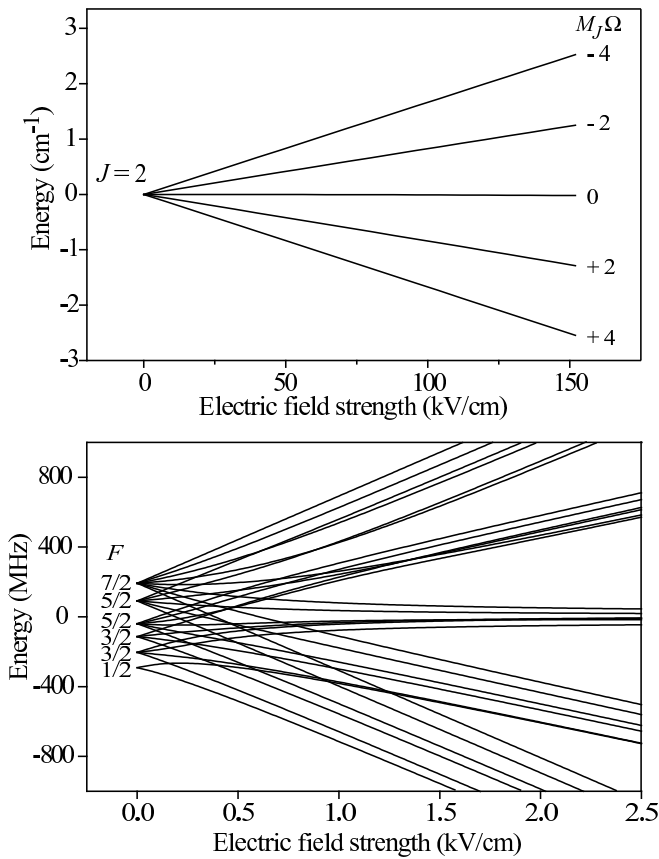}}
\caption{\label{fig:stark} The Stark shift of the metastable
$a\,^{1}\Delta(v=0,J=2)$ state. In the upper panel the energy
splitting (in cm$^{-1}$) in strong electric fields is shown, not
resolving the hyperfine structure. The lower panel shows the
energy splitting (in MHz) in low electric fields, including the
hyperfine structure of $^{14}$NH.}
\end{figure}

A molecular beam of NH molecules is formed by photo-dissociation
of about 1 vol~\% hydrogen azide (HN$_{3}$) seeded in Krypton by
focussing the 6 mJ output of a ArF excimer laser on a small quartz
capillary that is mounted on the tip of a pulsed supersonic valve.
Gaseous HN$_3$ is prepared by the reaction of sodium azide
(NaN$_3$) with an excess of lauric acid under vacuum. The reaction
starts at approximately 348 K and proceeds smoothly at 353-363 K.
The NH radicals are produced both in the $X\,^3\Sigma^-$ ground
state and in the metastable $a\,^1\Delta$ electronically excited
state. With Krypton as a carrier gas the molecular beam has a
forward velocity centered at about 450- 500 m/s with a full-width
half-maximum (FWHM) velocity spread of about 20 \%. The use of
Xenon as a carrier gas, which would result in a lower initial
velocity, is not possible since it is known to efficiently quench
the metastable NH molecules~\cite{Hack1989}. After the supersonic
expansion, most molecules in the $a\,^1\Delta$ electronic state
reside in the $v=0, J=2$ ro-vibrational ground state. The Stark
shift of this level is shown in the upper part of
Figure~\ref{fig:stark}. Only molecules in the low-field seeking
$M_J\Omega=-4$ component are decelerated and trapped in the
experiments.

On a smaller energy scale, in the lower part of
Figure~\ref{fig:stark}, the effect of the hyperfine structure of
$^{14}$NH on the Stark splitting becomes visible. Due to the
nuclear spin of both the $^{14}N$ and the $H$ nucleus, the
$a\,^1\Delta, J=2$ level is split into six hyperfine levels that
are labelled by the quantum number $F$. The $\Lambda$-doublet
splitting of the $J=2$ level is only 116 kHz~\cite{Ubachs1986},
and is not visible on this energy scale. The hyperfine splitting
is therefore (much) larger than the $\Lambda$-doublet splitting,
and the low- and high-field seeking $M_F$ components of the
different hyperfine states interact strongly. This unusual
situation is rather different from that in the molecules that have
been used in Stark deceleration experiments thus far, and one may
wonder (and worry) if this complex energy level structure could
result in a redistribution over the $M_J\Omega$ states during the
deceleration process.

Approximately 3~cm downstream from the nozzle orifice, the
molecular beam passes through a skimmer, and enters the
decelerator chamber. The molecules are focussed into the Stark
decelerator by a short hexapole. The Stark decelerator that is
installed in the trapping machine consists of 108~electric field
stages with a center to center distance of 11~mm. Each stage
consists of 2~parallel 6~mm diameter electrodes that are placed
around the molecular beam axis to provide a 4 $\times$ 4~mm$^2$
opening area for the molecular beam to pass. A voltage difference
of 40 kV is applied to the electrodes. At the end of the
decelerator an electrostatic quadrupole trap is mounted. The trap
loading procedure is similar to that used to trap
OH~\cite{VANDEMEERAKKER2005}. The Stark decelerator that is used
in the deceleration experiments consists of 104~electric field
stages. The electrode geometry of this decelerator is identical to
that of the other decelerator, except for an overall scaling
factor of 0.75, i.e., the electrodes have a 4.5~mm diameter, are
placed 8.25~mm apart, and provide a 3 $\times$ 3 mm$^2$ aperture.
However, the same voltage difference of 40~kV is applied to the
electrodes. A photograph of the latter decelerator is shown in the
upper left corner of Figure~\ref{fig:setup}. As can be seen from
this photograph, the last 8 electric field stages are mounted on
conically shaped rods to provide an optimal solid angle for
fluorescence collection.

The NH ($a\,^{1}\Delta$) radicals that exit the decelerator are
state-selectively detected using a far off-resonant Laser Induced
Fluorescence (LIF) scheme that is schematically shown in the inset
of Figure \ref{fig:setup}. The molecules are excited to the upper
$\Lambda$-doublet component of the $A\,^3\Pi_1, v=0, J=1$ level on
the $P_2(2)$ line of the spin-forbidden $A\,^{3}\Pi$ $\leftarrow$
$a\,^{1}\Delta$ transition around
584~nm~\cite{VANDEMEERAKKER2003}, and the subsequent fluorescence
to the $X\,^{3}\Sigma^{-}$ ground state around 336 nm is recorded.
Although the use of a spin-forbidden transition for LIF detection
of molecules is unconventional, this scheme is advantageous as it
allows almost background-free detection by blocking the
stray-light from the excitation laser with optical filters. The
584 nm light is generated by a narrow-band pulsed laser system. In
this laser the output of a frequency stabilized single mode ring
dye laser is amplified in a three stage pulsed dye amplifier
pumped by a frequency-doubled injection seeded Nd:Yag pump laser.
For the excitation, a 3~mm diameter laser beam with a pulse energy
of up to 25 mJ in a 5 ns duration pulse, with a bandwidth of
approximately 130 MHz, is used. This intensity is sufficient to
saturate even this weak spin-forbidden transition. In the
deceleration experiment, the laser beam excites the molecules just
behind the last electrodes of the decelerator. In the trapping
experiments, the laser beam is directed through holes in the
center trap electrode, and the fluorescence is collected through a
6~mm hole in the end-cap electrode. The fluorescence is imaged
with a lens onto a photomultiplier tube. Unless stated otherwise,
all electrodes are grounded 10~$\mu$s before the laser excites the
molecules.

\section{Results and discussion}

\subsection{Deceleration of NH ($a\,^1\Delta$) radicals}

\begin{figure}
    \resizebox{\linewidth}{!}{\includegraphics{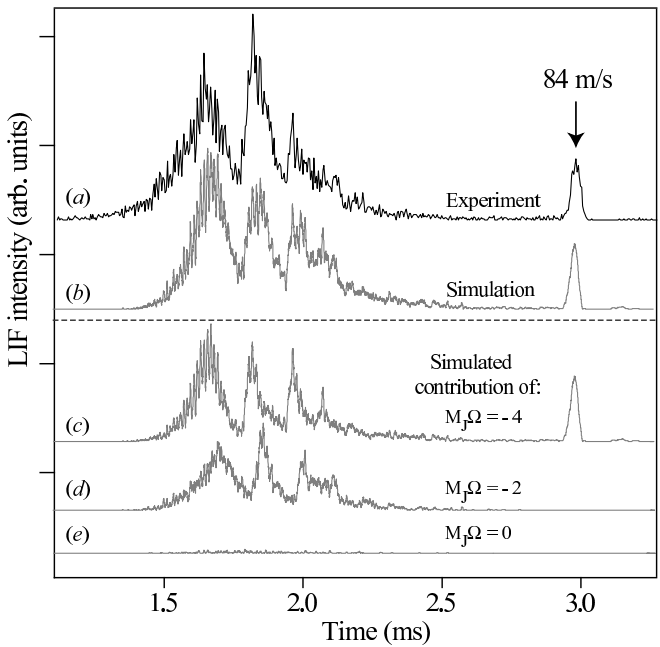}} \caption{\label{fig:TOFdeceleration} A measured time-of-flight (TOF) profile of metastable NH molecules observed in a deceleration experiment (\emph{a}), compared with the outcome of a numerical simulation (\emph{b}). The molecules are decelerated from 520 m/s to a final velocity of 84 m/s. The contribution of individual $M_J\Omega$ components to the simulated TOF-profile is shown in the lower part of the figure.}
    \end{figure}

A typical time-of-flight (TOF) profile of NH ($a\,^1\Delta, v=0,
J=2$) radicals that is observed in a deceleration experiment is
shown in curve (\emph{a}) of Figure \ref{fig:TOFdeceleration}. In
this experiment, a packet of molecules with an initial velocity of
520~m/s is selected, and the Stark decelerator is programmed to
extract 1.6 cm$^{-1}$ of kinetic energy from the packet in every
electric field stage.
The selected fraction of the molecular beam pulse that is
decelerated to a final velocity of 84~m/s arrives in the detection
region about 3 ms after production, and is split off from the
remaining part of the beam that is not decelerated. The TOF
profile that is obtained from a three dimensional trajectory
simulation of the experiment is shown in curve (\emph{b}), and is
seen to quantitatively reproduce the observed TOF profile. In the
simulations, the NH radicals are assumed to be produced in each of
the six $M_J\Omega$ components of the $J=2$ state with equal
probability. In the curves (\emph{c}), (\emph{d}), and (\emph{e})
the individual contributions of the $M_J\Omega=-4$, the
$M_J\Omega=-2$, and both $M_J\Omega=0$ components to the TOF
profile are shown, respectively. The decelerated packet of
molecules exclusively consists of molecules in the $J=2,
M_J\Omega=-4$ state, but molecules in the $M_J\Omega=-2$ state
significantly contribute to the signal of the non-decelerated part
of the beam. The $M_J\Omega=0$ components hardly contribute to the
signal as molecules in these components do not experience
transverse focussing forces as they progress through the
decelerator. Molecules in the high-field seeking $M_J\Omega=+2$
and $M_J\Omega=+4$ components are deflected from the molecular
beam axis and do not reach the exit of the Stark decelerator.

From these simulations, the $M_J\Omega$ composition of the
structured TOF profile of the non-decelerated part of the beam is
easily identified. This structure, and in particular also the
ratio between signal intensity of the decelerated and
non-decelerated part of the beam pulse, is well reproduced by the
simulations. This is a strong indication that there is no
significant redistribution over the various $M_J\Omega$ components
when switching from one high-voltage configuration to the next in
the Stark decelerator.

\subsection{Trapping of NH ($\bold{a\,^1\Delta}$) radicals}

When the molecules are brought to a standstill the electrostatic
trap is switched on. The details of the trap loading process are
published elsewhere~\cite{VANDEMEERAKKER2005}.
        \begin{figure}
        \resizebox{\linewidth}{!}{\includegraphics{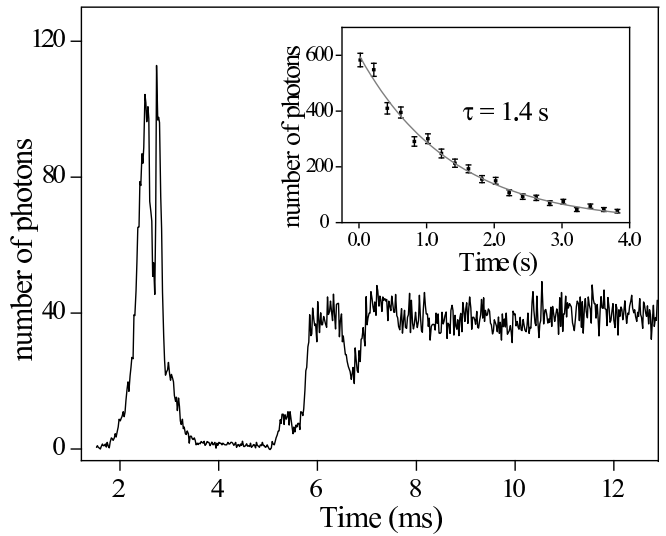}} \caption{\label{fig:trapping}The arrival of undecelerated NH molecules in the trap center is followed by the steady signal of trapped NH molecules. The molecules are brought to a standstill 6 ms after their production. At that moment the trap is switched on. In the inset the population of electrostatically trapped NH($a^{1}\Delta, v=0, J=2$) molecules is plotted as a function of time. The solid line is a single exponential fit to the data with an 1/e lifetime of 1.4 s.}
        \end{figure}
The intensity of the LIF signal of NH ($a\,^1\Delta,v=0, J=2$)
radicals, detected at the center of the trap, is shown as a
function of time in Figure~\ref{fig:trapping}. The trap is
switched on about 6 ms after the NH radicals are produced. After a
few initial oscillations a steady signal of the trapped molecules
remains. In the inset the signal of the trapped molecules is
plotted on a longer time scale. We have fitted a single
exponential decay to the data, resulting in a 1/$e$ trapping time
of 1.4 $\pm$ 0.1 s.

There are three main processes that limit the trapping time:
rotational excitation due to absorption of room-temperature
blackbody radiation, collisions with the background gas and the
spontaneous decay of the metastable state itself. In a recent
study, the effects of blackbody radiation have been quantified for
OH/OD, and tabulated for a variety of
molecules~\cite{Hoekstra:2007ly}. For NH($a\,^1\Delta,v=0,J=2$)
the loss rate due to optical pumping by room-temperature blackbody
radiation is 0.36 s$^{-1}$. The most abundant collision partner
for the trapped metastable NH radicals is Krypton, the carrier gas
used in these experiments. Unfortunately this loss factor can not
be quantified because the NH($a\,^1\Delta$)-Krypton collision
cross-section is unknown. The radiative lifetime of the
$a\,^1\Delta$ state of NH is not known accurately, but is expected
to be on the order of a few seconds (see~\cite{RINNENTHAL1999A}
and references therein). When we combine the measured trapping
time with the calculated loss rate due to blackbody radiation
pumping and when we assume no loss due to collisions with the
background gas, we find a lower limit for the radiative lifetime
of the NH($a\,^{1}\Delta, v=0,J=2$) state of 2.7 seconds.

The number of trapped molecules can be estimated from the
intensity of the detected fluorescence. The averaged number of
detected photons, after 200 ms of trapping, is 2.4 per excitation
laser pulse. From this the number of trapped
NH($a\,^1\Delta,v=0,J=2$) molecules is estimated to be about
$10^4$, corresponding to a density on the order of $10^6$
molecules cm$^{-3}$. This is about one order of magnitude less
than in our OH trapping experiments, reflecting the inferior beam
intensity of NH($a\,^1\Delta$) radicals. The production of
NH($a\,^1\Delta$) radicals in the beam can be improved when the
HN$_3$ molecules are photodissociated using the 266 nm light of a
Nd:Yag laser. In this scheme, the production of NH
($X\,^3\Sigma^-$) is spin forbidden, and the NH radicals are
exclusively produced in the $a\,^1\Delta$
state~\cite{McDonald1977}. Alternatively, NH ($a\,^1\Delta$)
molecules can be produced with a high quantum yield by
photodissociation of HNCO at 193 nm~\cite{Drozdoski1979}.

\subsection{Spectral profiles}

An important property of the trapped gas is the temperature.
Measuring the temperature, however, is notoriously difficult. The
temperature can in principle be determined by measuring the
spatial distribution of the molecular cloud, as the spatial and
velocity distribution of the molecules in the trap are coupled. In
trapping experiments with ND$_3$ molecules, that are detected
using a Resonance Enhanced Multi Photon\ Ionization (REMPI) scheme
with a focussed laser, the spatial distribution can be measured by
mechanically scanning the laser focus along a symmetry axis of the
trap~\cite{BETHLEM2002}. This strategy can only be implemented if
the detection laser beam is (much) smaller than the size of the
molecular cloud. When LIF is used as a detection scheme, the laser
beam can usually not be made small enough to measure the spatial
distribution this way. When the laser beam is comparable in size
to the molecular cloud, however, the temperature can be deduced
from the spectral profile of the molecules in the presence of the
trapping potential, a method that has, for instance, been applied
to magnetically trapped CaH molecules~\cite{WEINSTEIN1998}.
Unfortunately, if the excitation wavelength is in the ultraviolet
part of the spectrum, scattered photons can release electrons from
the trap electrode material. This can cause electrical breakdown
in electrostatic traps, and has thus far hampered the application
of this method to electrostatically trapped molecules.

\begin{figure}
\resizebox{\linewidth}{!}{\includegraphics{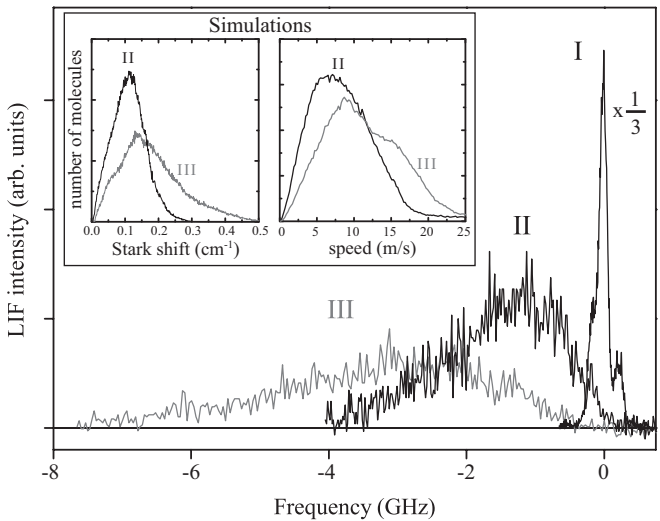}}
\caption{\label{fig:spectra}Spectral profiles obtained by scanning
the detection laser frequency over the highest frequency component
of the $P_2(2)$ transition, which excites the trapped NH molecules
from the $a^{1}\Delta$ to the $A^{3}\Pi$ state. The field-free
profile (I) is shown, scaled down by a factor 3, together with the
profiles obtained while the shallow (II) or deep (III) trapping
field is on. In the inset the result of numerical simulations is
given, showing the number of trapped molecules as a function of
their Stark shift and speed.}
\end{figure}
The detection scheme that is used in the present experiments does
not suffer from these limitations, as the 584~nm radiation can
safely be used to excite the molecules while the trap electrodes
are still on high voltage. In the main part of
Figure~\ref{fig:spectra} three different spectral profiles of the
$P_2(2)$ line are shown, all recorded after 16 ms of trapping. The
origin of the horizontal axis is set at the center of the
field-free line profile, indicated as profile (I). This line
profile is measured 10 $\mu$s after the trap electrodes are
switched off and grounded. This time interval is sufficiently
short that the molecules do not move significantly, while being
long enough to ensure field-free conditions. A near-Gaussian
lineshape with a FWHM of $\sim$ 130 MHz, mainly determined by the
bandwidth of the laser, is observed. Weak shoulders due to partly
resolved hyperfine structure and the intrinsic spectral structure
in the laser pulse can be recognized in this profile.

Also shown are two broadened line profiles, (II) and (III),
observed when the molecules are excited in the presence of the
trapping potential. In the electric field of the trap, the trapped
molecules in the $a\,^1\Delta,v=0,J=2$ level and the $A\,^3\Pi_1,
v'=0,J'=1$ level both experience a positive Stark shift. Since the
Stark shift for molecules trapped in the $a^{1}\Delta$ state is
considerably larger than the Stark shift in the excited
$A^{3}\Pi_{1}$ state, a broadened profile of the $P_2(2)$ line
results that is shifted to lower frequencies. The difference
between profile (II) and (III) reflects the difference in the
depth and slope of the trapping potentials that are being used;
the deceleration and trap loading process is identical in both
cases. Profile (II) is recorded when the ring electrode of the
trap is grounded, whereas a voltage of +15 kV is applied to this
electrode to record profile (III). The endcaps of the
electrostatic trap are at a potential of -15 kV in both cases. The
resulting trap potentials therefore differ in slope and depth by a
factor of two, and are referred to hereafter as 'shallow' (II) and
'deep' (III).

The laser excitation and fluorescence detection in the experiment
is done through holes in the trap electrodes, inevitably
restraining access to the trapping volume. For a direct
determination of the temperature of the trapped molecules from the
measured line profiles a complicated instrument function would
have to be taken into account; the laser position, its spatial
intensity profile and the spatial sensitivity of the detection
system have to be accurately known. Therefore, a detailed analysis
of the line profiles intrinsically has many uncertainties.
However, information on the energy and the position of the
molecules in the trap can also be obtained in a straightforward
manner by comparing the experimental line profiles with each
other. In such a comparison, complications related to details of
the excitation- and detection-processes cancel out.

The integrated signal of the three line profiles is, within
experimental accuracy, the same. Moreover, if profile (II) is
stretched out in frequency by a factor of two, and reduced in
intensity by a factor of two, it is virtually identical to profile
(III). This alone indicates that the spatial distribution of the
trapped molecules in the shallow and deep potential are the same.
This in turn implies that the initial kinetic energy of the
molecules, at the moment that the trap is switched on, is small
relative to the depth of the trapping potentials. The energy
distribution (temperature) is therefore mainly determined by the
trapping potential, and is about a factor of two different for the
shallow and the deep trap.

This conclusion is supported by numerical simulations of the
deceleration and trapping process. The inset in
Figure~\ref{fig:spectra} shows the results of these simulations,
indicating the number of trapped molecules as a function of their
Stark shift and speed, for both trapping geometries. The molecules
in the shallow trap potential~(II) have a smaller Stark shift and
a narrower speed distribution than the molecules trapped in the
deep trap potential~(III). The simulated speed distribution can be
characterized by a temperature, which is on the order of 60 mK and
100 mK for the molecules trapped in the shallow and deep trap
potential, respectively.

\section{conclusions}

A pulsed molecular beam of NH radicals in the metastable
$a^{1}\Delta$ state has been produced, Stark decelerated, and
loaded into an electrostatic trap. The density of the trapped
sample is on the order of $10^{6}$ cm$^{-3}$, which can be further
enhanced by implementing more efficient production schemes. The
1/$e$ trapping lifetime is determined as 1.4$\pm$0.1 s, which
yields a lower limit of 2.7 s for the radiative lifetime of the
$a\,^1\Delta,v=0,J=2$ state. The spatial distribution of the
trapped cloud is probed by performing spectroscopy on the
molecules in the presence of the trapping field. This experiment
is an important step towards the accumulation of multiple packets
of Stark-decelerated molecules in a magnetic trap, which is a
promising route towards higher densities of cold molecules.
Modifications to the machine that will enable the transfer of the
NH radicals into a magnetic trap are currently being implemented.

\section{Acknowledgements}
We acknowledge the experimental support from Henrik Haak and Sandy
Gewinner, and the theoretical support from Boris Sartakov. M.M. is
grateful to the Academy of Finland for financial support.

\bibliography{references}
\end{document}